# Radar pulse correlation from uniformly translating targets

Timothy J. Garner[a]* and Akhlesh Lakhtakia[b]

[a]*Army Research Laboratory, Adelphi, MD, USA;* [b]*Department of Engineering Science and Mechanics, Pennsylvania State University, University Park, PA, USA*

*timothy.j.garner.civ@mail.mil

# Radar pulse correlation from uniformly translating targets

Radars often use correlation of received signals with transmitted signals to identify targets. However, when a target translates at a high uniform speed, the correlation of the transmitted and received signals depends strongly on the target's velocity, even after the two-way Doppler shift has been removed. This is true whether the target is made of highly dispersive material or not.

**Introduction**

Radar identification is done by pinging a target with a time-limited signal modulating the amplitude of a carrier wave. Very often, the return ping is recorded in the backscattered direction. The time-of-flight, which tells the distance to the target, may be determined by correlating a time-shifted copy of the transmitted pulse with the received pulse [1]. However, if the target is flying rapidly, the return pulse will be affected by the velocity of the target, even after removing the two-way Doppler shift from the return pulse. The electromagnetic pulse backscattered by a uniformly translating target may differ greatly from the incident pulse, depending on the velocity of the target as well as on its shape, orientation, and composition.

In this Letter, we compute the Pearson correlation coefficient between the incident and backscattered signals from three uniformly translating targets: (i) a silicon-carbide rod, (ii) a silicon-carbide disk, and (iii) a metallic model spacecraft. We examine the relationships between the correlation coefficient and the target's velocity.

**Methods and theory**

Consider a target moving with constant velocity $v'$ in the radar's reference frame $K'$. Let $K$ be a co-moving inertial frame chosen such that the target is affixed to the origin of $K$. The origins and axes of $K'$ and $K$ are chosen to align at time $t' = t = 0$. Primed and unprimed space-time coordinates are used in $K'$ and $K$, respectively. Vectors are denoted with bold type, and unit vectors are decorated with carets. The speed of light in free space is denoted by $c$.

In $K'$, an electromagnetic signal propagating in the $+z'$ direction with its electric field aligned parallel to the $x'$ axis is incident on the target. The pulse transmitted by the radar is denoted by $f(t')$, and the pulse received by the radar with the two-way Doppler shift [2] removed is denoted by $g(t')$. A complete description of the method used to compute the received pulse is given elsewhere [3].

We computed the Pearson correlation coefficient $\rho$ of the time-sampled versions of the incident and backscattered signals in $K'$ after modifying the backscattered signal to remove the two-way Doppler shift. The Pearson correlation coefficient is defined as [4]

$$\rho_{fg} = \frac{\sum_{n=1}^{N}(f_n - \mu_f)(g_n - \mu_g)}{\sqrt{\sum_{n=1}^{N}(f_n - \mu_f)^2}\sqrt{\sum_{n=1}^{N}(g_n - \mu_g)^2}}, \qquad (1)$$

where the $f_n$ and $g_n$ are samples of the incident and backscattered signals, respectively, evaluated at $t'_n = nt'_s$, and $t'_s$ is the sampling interval. For each pair of incident and scattered signals, we time-shifted the incident signal relative to the backscattered signal to find $\rho_{\max}$, the maximum absolute value of the correlation $\rho_{fg}$ between the two signals. $\mu_f$ and $\mu_g$ are the sample means of the incident and backscattered signals.

**Numerical results**

We used previously published backscattered signals from a uniformly translating silicon-carbide rod, a uniformly translating disk, and a uniformly translating model spacecraft [5]. The three targets are depicted in Figure 1. For all three targets, the incident signals were plane waves that were amplitude modulated by a Gaussian pulse. For the silicon-carbide rod and disk, the carrier frequency $\nu'$ of the plane wave was 20 THz and the width parameter $\sigma'$ of the Gaussian pulse was 50 fs. For the model spacecraft, $\nu' = 20$ GHz and $\sigma' = 50$ ns. We used the *corrcoef* function from the NumPy library [6] for Python to compute the Pearson correlation coefficient.

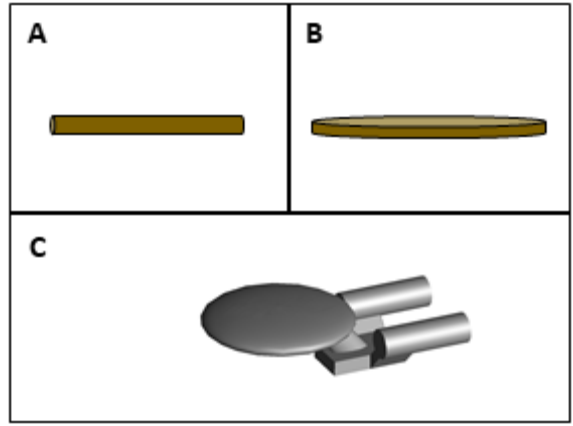

Figure 1. Silicon-carbide rod (A), silicon-carbide disk (B), and metallic model spacecraft (C) used as targets.

*Silicon-carbide rod*

Table 1 shows the maximum correlation between the incident and backscattered signals for the silicon-carbide rod. The unit vector $\hat{n}$ points along the axis of the rod. The rod is 10 $\mu$m long and has a diameter of 1 $\mu$m in $K$. Note that, in general, $\rho_{max}$ is greatest when the rod is aligned with the $x$ axis, which is also the direction of the incident electric field in $K'$; in contrast, $\rho_{max}$ is lowest when the rod is aligned with the $z$ axis.

| Orientation ($\hat{n}$) | Velocity ($v'$) | Maximum correlation ($\rho_{max}$) |
|---|---|---|
| $\hat{x}$ | $-0.2c\hat{z}'$ | 0.8217 |
| $\hat{x}$ | $0.2c\hat{x}'$ | 0.9373 |
| $\hat{x}$ | $0.2c\hat{y}'$ | 0.5401 |
| $\hat{x}$ | $0.2c\hat{z}'$ | 0.8812 |
| $\hat{x}$ | 0 | 0.9383 |
| $\hat{y}$ | $-0.2c\hat{z}'$ | 0.7060 |
| $\hat{y}$ | $0.2c\hat{x}'$ | 0.7891 |
| $\hat{y}$ | $0.2c\hat{y}'$ | 0.5581 |
| $\hat{y}$ | $0.2c\hat{z}'$ | 0.9757 |
| $\hat{y}$ | 0 | 0.8631 |
| $\hat{z}$ | $-0.2c\hat{z}'$ | 0.2978 |
| $\hat{z}$ | $0.2c\hat{x}'$ | 0.5714 |
| $\hat{z}$ | $0.2c\hat{y}'$ | 0.2805 |
| $\hat{z}$ | $0.2c\hat{z}'$ | 0.7737 |
| $\hat{z}$ | 0 | 0.5875 |

*Silicon-carbide disk*

Table 2 shows the maximum correlation between incident and backscattered signals from a silicon-carbide disk. The diameter of the disk is 20 $\mu$m and the thickness is 1 $\mu$m in $K$. The orientation of the disk in $K$ is specified by the unit vector $\hat{n}$ that is aligned with the disk's axis of symmetry in $K$. The maximum correlation $\rho_{max}$ is generally highest for $\hat{n} = \hat{z}$, which has the face of the disk aligned perpendicular to the direction of propagation of the incident signal when the velocity is aligned with the z axis.

| Orientation ($\hat{n}$) | Velocity ($v'$) | Peak correlation ($\rho_{max}$) |
|---|---|---|
| $\hat{x}$ | $-0.2c\hat{z}'$ | 0.4043 |
| $\hat{x}$ | $0.2c\hat{x}'$ | 0.3993 |
| $\hat{x}$ | $0.2c\hat{y}'$ | 0.0872 |
| $\hat{x}$ | $0.2c\hat{z}'$ | 0.6611 |
| $\hat{x}$ | 0 | 0.3351 |
| $\hat{y}$ | $-0.2c\hat{z}'$ | 0.4425 |
| $\hat{y}$ | $0.2c\hat{x}'$ | 0.5287 |
| $\hat{y}$ | $0.2c\hat{y}'$ | 0.0003 |
| $\hat{y}$ | $0.2c\hat{z}'$ | 0.6071 |
| $\hat{y}$ | 0 | 0.4630 |
| $\hat{z}$ | $-0.2c\hat{z}'$ | 0.9332 |
| $\hat{z}$ | $0.2c\hat{x}'$ | 0.9199 |
| $\hat{z}$ | $0.2c\hat{y}'$ | 0.0078 |
| $\hat{z}$ | $0.2c\hat{z}'$ | 0.9829 |
| $\hat{z}$ | 0 | 0.9412 |

*Metallic model spacecraft*

We also computed the maximum correlation between the incident and backscattered signals from a uniformly translating model spacecraft made of an infinitely conducting material [5]. The model spacecraft is 8.4 cm long and 5.0 cm wide. The correlation is dependent on velocity, although a simple relationship between the two could not be found.

| Orientation (facing toward / away from source) | Velocity ($v$) | Peak correlation ($\rho_{max}$) |
|---|---|---|
| Toward | $-0.9c\hat{z}'$ | 0.8223 |
| Toward | $-0.5c\hat{z}'$ | 0.3868 |
| Toward | $-0.2c\hat{z}'$ | 0.8916 |
| Toward | $-0.1c\hat{z}'$ | 0.9489 |
| Toward | $-0.05c\hat{z}'$ | 0.9337 |
| Toward | 0 | 0.8963 |
| Away | 0 | 0.8576 |
| Away | $0.05c\hat{z}'$ | 0.8328 |
| Away | $0.1c\hat{z}'$ | 0.7947 |
| Away | $0.2c\hat{z}'$ | 0.8391 |
| Away | $0.5c\hat{z}'$ | 0.6965 |
| Away | $0.9c\hat{z}'$ | 0.6440 |
| Away | $0.99c\hat{z}'$ | 0.8663 |

**Concluding remarks**

The correlation between the incident signal and signal backscattered by a uniformly translating object is strongly dependent on the magnitude and direction of velocity, even when the two-way Doppler shift is removed from the backscattered signal. This is true both for targets made of a dispersive material such as silicon carbide and for targets made of an idealized (i.e., perfectly conducting) metal. In the case of the spacecraft model, the correlation is highly affected by the complexity of shape. Radar installations may need to adjust the carrier frequency or the bandwidth of the transmitted pulses to achieve sufficiently high correlation between the incident and returned pulses for foolproof detection. It may be necessary to use a trial-and-error approach to determine the appropriate pulse and carrier frequency for uniformly translating objects of a specific type.

**References**

[1] Kailath T. Correlation detection of signals perturbed by a random channel. IRE Transactions on Information Theory 1960;6(3):361-366.

[2] Van Bladel J. Relativity and Engineering. Berlin: Springer; 1984. Section 3.13.

[3] Garner TJ, Lakhtakia A, Breakall JK, et al. Time-domain electromagnetic scattering by a sphere in uniform translational motion. Journal of the Optical Society of America A 2017;34(2):270-279.

[4] Rodgers JL and Nicewander WA. Thirteen ways to look at the correlation coefficient. The American Statistician 1988;42(1):59-66.

[5] Garner TJ, Lakhtakia A, Breakall JK, et al. Electromagnetic pulse scattering by a spacecraft nearing light speed. Applied Optics 2017;56(22):6206-6213.

[6] numpy.org [Internet]. Austin, TX, USA: NumFOCUS; [cited 2007 June 9]. Available from: https://numpy.org/